\documentclass{article}

\usepackage[utf8]{inputenc}
\usepackage{amsmath}
\usepackage{amssymb}
\usepackage{amsthm}
\usepackage{graphicx}
\usepackage{natbib}
\usepackage{url}
\usepackage{hyperref}
\usepackage{pifont}

\usepackage{footnote}
\makesavenoteenv{tabular} 

\newcommand{\cmark}{\ding{51}}
\newcommand{\xmark}{\ding{55}}
\newcommand{\pnoise}{p_\text{noise}}

\title{Balancing Cooperativeness and Adaptiveness in the (Noisy) Iterated Prisoner's Dilemma}
\author{Adrian Hutter\\\texttt{ahut@google.com}}
\date{\today}

\begin{document}

\maketitle

\begin{abstract}

Ever since Axelrod’s seminal work, tournaments served as the main benchmark for evaluating strategies in the Iterated Prisoner's Dilemma (IPD).
In this work, we first introduce a strategy for the IPD which outperforms previous tournament champions when evaluated against the 239 strategies in the {\fontfamily{qcr}\selectfont Axelrod} library, 
at noise levels in the IPD ranging from 0\% to 10\%.
The basic idea behind our strategy is to start playing a version of tit-for-tat which forgives unprovoked defections if their rate is not significantly above the noise level, 
while building a (memory-1) model of the opponent; 
then switch to a strategy which is optimally adapted to the model of the opponent.
We then argue that the above strategy (like other prominent strategies) lacks a couple of desirable properties which are not well tested for by tournaments, but which will be relevant in other contexts:
we want our strategy to be \emph{self-cooperating}, i.e., cooperate with a clone with high probability, even at high noise levels; 
and we want it to be \emph{cooperation-inducing}, i.e., optimal play against it should entail cooperating with high probability.
We show that we can guarantee these properties, at a modest cost in tournament performance, 
by reverting from the strategy adapted to the opponent to the forgiving tit-for-tat strategy under suitable conditions.

\end{abstract}

\section{Introduction} 

Direct reciprocity is one of the main mechanisms explaining the emergence of cooperation between self-interested individuals \citep{nowak06}, 
and the Iterated Prisoner's Dilemma (IPD) is the primary model used for the study of direct reciprocity. 
Famously, tit-for-tat (TFT) won both of Robert Axelrod's seminal IPD tournaments \citep{axelrod84}.
In any realistic situation resembling an IPD, intended actions can be executed imperfectly or otherwise have unintended consequences.
The simplest way of modelling this in the IPD is to assume that each action has the opposite of the intended effect with some probability $\pnoise$.
It was soon recognized that the noisy IPD introduces fundamentally new challenges. 
In particular, for TFT a single noise event ends mutual cooperation (at least till the next noise event occurs), 
leading to alternating rounds of C/D, D/C instead \citep{axelrod81}.
This motivated the development of alternatives to TFT which are able to cooperate more robustly in the presence of noise, 
including tit-for-two-tats \citep{axelrod84}, Generous TFT \citep{molander85}, Pavlov (\textit{aka.}\ Win-Stay Lose-Shift) \citep{nowak93}, and Contrite TFT \citep{wu95}.
\citet{wu95} evaluated the latter three strategies against all 63 strategies submitted to Axelrod's second tournament \citep{axelrod84} (using $\pnoise=1\%$), 
finding Generous and Contrite TFT to perform strongly, and Pavlov to perform poorly.

In celebration of the 20th anniversary of Axelrod's seminal work, \citet{kendall07} organized two IPD tournaments (both noise-free and noisy, with $\pnoise=10\%$) in 2004 and 2005.
We briefly describe two of the best-performing strategies, both of which are designed specifically to account for noise.
Omega TFT \citep{slany07} modifies TFT in two ways: if it detects a deadlock loop (alternating rounds of C/D and D/C, which are characteristic for TFT) it attempts to break the loop by playing C twice; 
and it reverts to playing D for the remainder of the game if a measure of randomness of the opponent's play exceeds a threshold.
DBS \citep{au07} models the opponent as a memory-1 strategy.\footnote{I.e., the opponent is described by the probabilities of playing C after the 4 possible states $\{\text{CC}, \text{CD}, \text{DC}, \text{DD}\}$.}
It attempts to describe the opponent using deterministic rules, and ignores occasional violations of these rules (which might be due to noise). It then optimizes its move against the model of the opponent using tree search to depth 5.

More recently, \citet{harper17} trained various model types (lookup tables, artificial neural networks, finite state machines, hidden Markov models),
in both noisy and noise-free environments, against a large zoo of strategies from previous literature.
They then ran a noisy ($\pnoise=5\%$) and a noise-free tournament of 176 strategies, including the trained strategies and all strategies mentioned so far.
Remarkably, DBS was the best-performing human-designed strategy in both tournaments, ranking 1st in the noisy tournament and 12th in the noise-free tournament.
Omega TFT ranked not far behind DBS (8th in the noisy tournament and 15th in the noisy one).
The first ranked strategy in the noise-free tournament was EvolvedLookerUp2\_2\_2, a deterministic lookup table, 
which bases its decision on the first 2 actions of the opponent and the past 2 actions of itself and the opponent.
All strategies participating in these tournaments (and more) are available in the {\fontfamily{qcr}\selectfont Axelrod} library \citep{knight16}, 
which we will use for evaluation purposes in the present work.

The goal of the present work is two-fold.
Firstly, we want to create a strategy which is both able to robustly cooperate (like the forgiving variants of TFT) and highly adaptive (by building a simple model of the opponent and responding optimally to that, like DBS).
These goals cannot be pursued at the same time, so we need rules for switching between the cooperative and the adaptive state. 
Sec.~\ref{sec:cooperateadapt} introduces such a strategy and evaluates it against all strategies in the {\fontfamily{qcr}\selectfont Axelrod} library.
Secondly, we discuss properties which are desirable for a strategy to have, beyond doing well in tournaments.
In Sec.~\ref{sec:desiderata}, we argue that a strategy should ideally be \emph{self-cooperating} (cooperate with a clone) and \emph{cooperation-inducing} (incentivize the opponent to cooperate).
We then show in Sec.~\ref{sec:car} how to make the previously introduced strategy self-cooperating and cooperation-inducing by introducing additional rules for switching between the cooperative and the adaptive state.

\section{Cooperate, then adapt to the opponent}\label{sec:cooperateadapt}

\begin{figure}\centering
\includegraphics[width=1.0\textwidth]{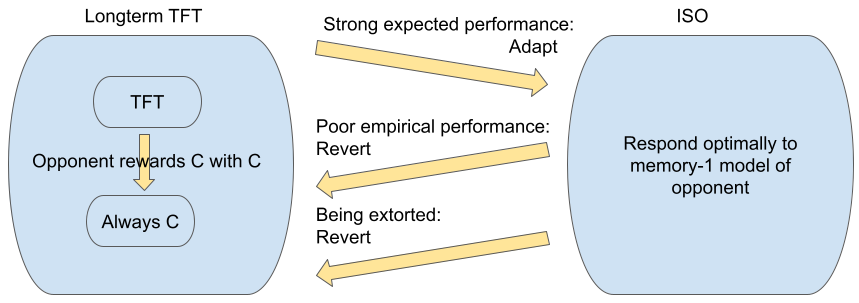}
\caption{The strategy introduced in this work comprises two sub-strategies, Longterm TFT (focused on allowing robust cooperation in the presence of noise) and ISO (focused on playing optimally against simple opponents), 
together with rules for switching between these. Sec.~\ref{sec:cooperateadapt} introduces the two sub-strategies and the rule for switching from Longterm TFT to ISO; 
Sec.~\ref{sec:car} introduces two rules for switching back, which are motivated in Sec.~\ref{sec:desiderata}.}
\label{fig:adapt_revert}
\end{figure}

The strategy introduced in this section, which we call \emph{CooperateISO}, comprises two sub-strategies: 
a highly cooperative one, and a purely adaptive one, together with a rule for switching from the former to the latter.
The cooperative strategy, \emph{Longterm TFT} focuses on allowing robust cooperation in the presence of noise without being exploitable.
The adaptive strategy, which we call \emph{ISO} (Infinite Sum Optimizer), 
builds a memory-1 model of the opponent and responds optimally to that.
The basic idea is illustrated in Fig.~\ref{fig:adapt_revert}, and motivated and explained in more detail in the following subsections.

In the following, we use the standard notation for referring to the possible payoffs in the PD, $T>R>P>S$, and use the conventional values $T=5$, $R=3$, $P=1$, and $S=0$.

\subsection{Longterm TFT: a maximally forgiving version of TFT}

Multiple existing strategies feature mechanisms for dealing with the effects of noise (e.g.\ \citet{molander85,nowak93,wu95,slany07,au07}).
In this subsection, we introduce a strategy which, like Generous TFT \citep{molander85}, is designed to be maximally forgiving (of defections which can be due to noise) while still incentivizing the opponent to cooperate.
However, while Generous TFT is a memory-1 strategy, our strategy takes (aggregate statistics of) the entire history of play into account; we thus call this strategy \emph{Longterm TFT}.

Longterm TFT starts by playing standard TFT. It switches to always cooperating if the history of play is compatible (given the noise model) with the opponent always rewarding cooperation with cooperation.
More precisely, let $N_C$ be the number of times our agent has cooperated so far, not taking our agent's most recenct action into account (since the opponent has not yet reacted to that).
(Note that we care about the \emph{actual} action of our agent here, taking the effects of noise into account, not the \emph{intended} action.)
Let $N_{C\rightarrow D}$ be the number of times the opponent defected after our agent cooperated. Define the statistic
\begin{align}\label{eq:z}
z = \frac{N_{C\rightarrow D} - \pnoise N_C}{\max\{1, \sqrt{\pnoise(1-\pnoise)N_C}\}}\ .
\end{align}

Under the null hypothesis that the opponent always rewards cooperation with cooperation, $N_{C\rightarrow D}$ is binomially distributed with mean $\pnoise N_C$ and variance $\pnoise(1-\pnoise)N_C$.
As a consequence of the Central Limit Theorem, $z$ will thus follow a standard normal distribution under the null hypothesis, if $\pnoise(1-\pnoise)N_C\ge1$ and $N_C$ is ``large enough''.
The statistic $z$ thus quantifies how confidently we can reject the null hypothesis that the opponent always rewards cooperation with cooperation.

Longterm TFT always cooperates if $N_C\ge5$ and $z<2$.\footnote{No attempt was made to optimize the precise threshold values, and similar thresholds in the rest of this work.}
For $\pnoise N_C<1$, i.e., if we expect less than one unprovoked defection under the null hypothesis,
the denominator in Eq.~(\ref{eq:z}) becomes $1$, and so Longterm TFT forgives at most 2 unprovoked defections.

During evaluation, we will assume that $\pnoise$ is known at the beginning of the IPD.
If it is not known, it has to be estimated from agreement between intended and actual actions. 
Since Longterm TFT always plays identically to TFT while $N_C<5$, it does not need an estimate of $\pnoise$ during the first few rounds of the IPD.

By forgiving defections which are not significantly more frequent than one would expect based on the noise model, 
Longterm TFT prevents unnecessary chains of retaliation when playing against a cooperative but provocable strategy like TFT and its variations.
It is thus able to robustly cooperate with such strategies (including itself) even in the presence of strong noise.

By being so forgiving, Longterm TFT invites occasional deliberate defections from the opponent.
However, the number of defections which are forgiven only grows like $O(\sqrt{N_C})$,
and so the \emph{rate} of defections which are forgiven goes to zero like $O(1/\sqrt{N_C})$.
The ``free'' defections which Longterm TFT forgives are thus irrelevant in the steady state.
In the steady state, Longterm TFT always cooperates against all strategies which are at least as cooperative as TFT: 
they may retaliate against Longterm TFT's (accidental) defections, but have to reward cooperation with cooperation.

\subsection{ISO: playing optimally against memory-1 opponents}\label{sec:iso}

Many of the most prominent strategies in the IPD are memory-1, i.e., 
their probability of cooperating on the next move only depends on the actions played by both players in the previous round.
Such strategies are thus described by four probabilities.\footnote{
Plus a fifth probability describing the probability of cooperating in the very first round of the IPD, which is irrelevant for our purposes.}
DBS \citep{au07} builds a memory-1 model of the opponent and then optimizes its own actions against that model.
The strategy described in this subsection, \emph{ISO} (Infinite Sum Optimizer), takes inspiration from DBS' remarkably strong performance in previous tournaments. 
ISO is both a simplification and a refinement of DBS.

DBS follows a complex algorithm for building its opponent model, trying to establish deterministic rules (such as ``the opponent always cooperates after mutual cooperation'') while taking possible noise events into account.
By contrast, ISO's opponent model is simply given by the (discounted) average rate of cooperation of the opponent in the four different states.
We use a discount factor $\gamma_\text{past}=0.99$ in order to react more quickly to changes in the opponent's strategy than if a simple average were taken.
In order to make the averages well-defined from the beginning, we assume that we have seen the opponent play the action dictated by TFT once for each of the four states.
We also clamp all empirical rates of cooperation to the interval $[\pnoise,1-\pnoise]$, since values outside of that interval are not feasible under the noise model.
Let $\vec{p}_\text{opp}^{\,(n)}$ denote our model of the opponent obtained this way,
i.e., the 4-vector of discounted average rates of cooperation in the four different states $(CC, CD, DC, DD)$.
We use the superscript $(n)$ to denote that these rates of cooperation already include the effects of noise, 
and so might be different from the \emph{intended} rates of cooperation.

While DBS uses tree search to depth 5 in order to optimize its action against the opponent model,
ISO optimizes the exact discounted sum of future payoffs.
Let $\vec{p}_\text{self}$ denote the 4-vector of probabilities of cooperation of our agent, which we hope to optimize against $\vec{p}_\text{opp}^{\,(n)}$.
In order to calculate how well a candidate $\vec{p}_\text{self}$ performs against a given $\vec{p}_\text{opp}^{\,(n)}$, 
we first calculate $\vec{p}_\text{self}^{\,(n)}$ by taking the effects of noise into account, i.e., 
applying $p\mapsto (1-\pnoise)p+\pnoise(1-p)$ to each entry in $\vec{p}_\text{self}$.
Iterated play between two memory-1 strategies (ISO and the model of the opponent) induces a Markov process of order 1.
We can then calculate the 4x4 transition matrix of this Markov process,
\begin{align}
T = \left(\vec{p}_\text{self}^{\,(n)}\odot\vec{p}_\text{opp}^{\,(n)},
\vec{p}_\text{self}^{\,(n)}\odot(1-\vec{p}_\text{opp}^{\,(n)}),
(1-\vec{p}_\text{self}^{\,(n)})\odot\vec{p}_\text{opp}^{\,(n)},
(1-\vec{p}_\text{self}^{\,(n)})\odot(1-\vec{p}_\text{opp}^{\,(n)})\right)
\end{align}
where $\odot$ denotes point-wise multiplication.
Let $\vec{u}=(R, S, T, P)$ denote the vector of possible payoffs, $\vec{s}_0$ a one-hot vector indicating the current state, and $\gamma_\text{future}<1$ a discount factor.
The average expected discounted payoff per round for $\vec{p}_\text{self}$ is then given by
\begin{align}\label{eq:U}
U(\vec{p}_\text{self}) &= \vec{s}_0^T\left(\sum_{k=1}^\infty \gamma_\text{future}^kT^k\right)\vec{u} / \left(\sum_{k=1}^\infty \gamma_\text{future}^k\right) \nonumber\\
&= (1-\gamma_\text{future})\vec{s}_0^T T\left(1-\gamma_\text{future}T\right)^{-1}\vec{u}\ .
\end{align}

If the IPD has a fixed ending probability $p_\text{end}$ per step, we can choose $\gamma_\text{future}=1-p_\text{end}$.
However, in the rest of this work, we will consider IPDs of fixed length, while assuming that the length was not known to the participants beforehand. 
We will use $\gamma_\text{future}=0.99$ in the rest of this work.

In order to find $\vec{p}_\text{self}$ which optimizes $U(\vec{p}_\text{self})$, we use the {\fontfamily{qcr}\selectfont Adam} optimizer \citep{kingma14} 
with a learning rate of 0.1 for 50 steps, using $(\frac{1}{2},\frac{1}{2},\frac{1}{2},\frac{1}{2})$ as the starting point of the optimization.

A conceptually similar strategy, IP$_0$, was introduced by \citet{lee15}. 
IP$_0$ optimizes payoffs in the stationary state against the memory-1 model of the opponent,
which is equivalent to Eq.~(\ref{eq:U}) in the limit $\gamma_\text{future}\rightarrow1^{-}$.

\subsection{CooperateISO: start cooperatively, then adapt}

Finally, \emph{CooperateISO} combines the two sub-strategies introduced so far.
It starts by playing Longterm TFT, and then switches to playing ISO if it has collected sufficient data about the opponent's behavior (which gets expressed in the memory-1 model)
and responding optimally to the opponent model promises higher payoffs than Longterm TFT has empirically achieved.

Note the trade-off involved in deciding when to start playing ISO.
On the one hand, playing Longterm TFT too long can be wasteful if a clearly better response to the opponent exists.
This includes opponents who do not retaliate against defections, but also highly defective strategies, against which Longterm TFT sub-optimally rewards each noise-induced cooperation.
On the other hand, switching to ISO is risky because it can destroy a cooperative relationship without achieving anything better if our model of the opponent is inaccurate.
This can happen because we have not seen the opponent react in all 4 possible states sufficiently many times,
because the opponent's empirically observed behavior is actually untypical for them because of noise events,
or because the opponent is simply not well described by a memory-1 model.

Note that in the presence of significant noise, maintaining cooperation with unforgiving strategies like Grim Trigger \citep{banks90} is hopeless irrespective of our behavior.
This suggests that the threshold for switching to ISO should be lower in the presence of high noise.
However, for simplicity we will use the same simple criterion for switching to ISO irrespective of the noise level.

Formally, let $N_c$ be the number of rounds played so far (while playing Longterm TFT), $U_c$ the average payoff per round achieved so far, $\sigma_c$ the corresponding standard deviation, 
and $U_a$ the expected average discounted payoff per round (from Eq.~(\ref{eq:U})) when playing ISO.
We switch from playing Longterm TFT to playing ISO if all of the following conditions are met: $N_c \ge 10$ (require a minimum of data about the opponent), 
$U_a - U_c > 2\sigma_c / \sqrt{N_c}$ (the expected gain from adapting is significant relative to the noise in the historical payoffs), and 
$U_a - U_c > 0.05(R-P)$ (the expected gain from adapting is non-negligible compared to possible payoff differences).

\subsection{Evaluation}

\begin{figure}\centering
\includegraphics[width=1.0\textwidth]{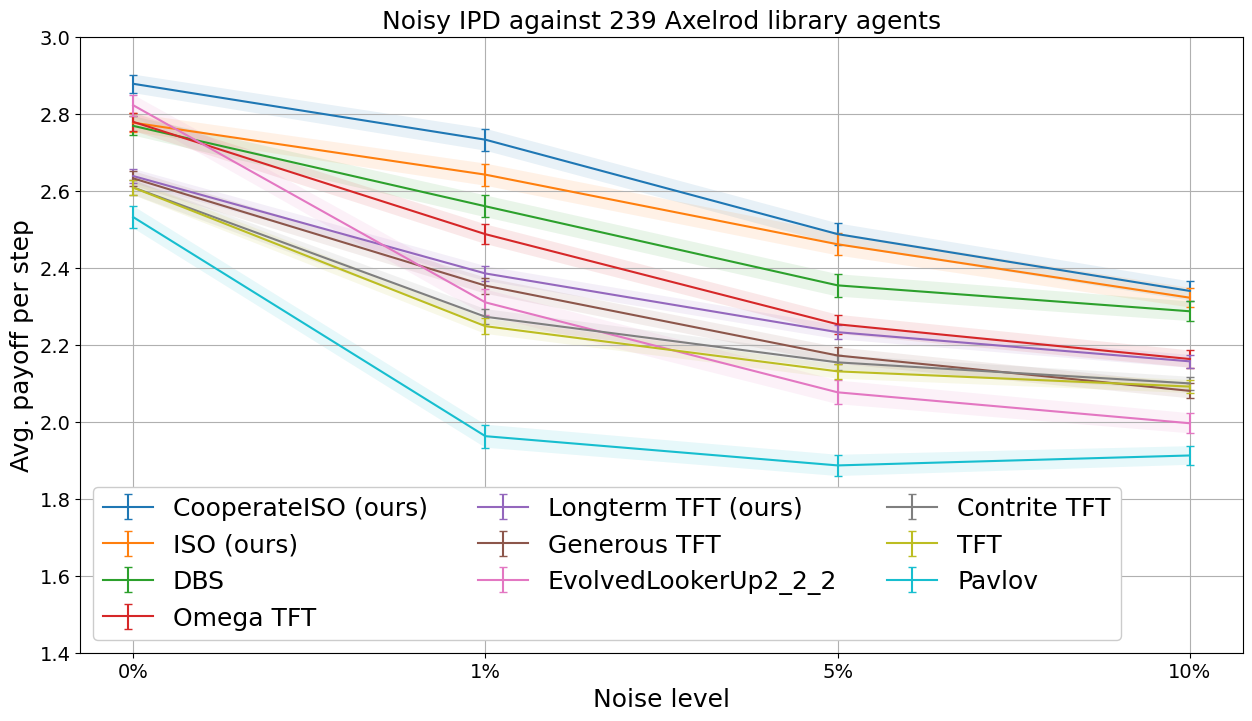}
\caption{Evaluation of the strategies introduced in Sec.~\ref{sec:cooperateadapt} (CooperateISO and its sub-strategies Longterm TFT and ISO) against the 239 strategies in the {\fontfamily{qcr}\selectfont Axelrod} library, evaluated at noise levels of $0\%$, $1\%$, $5\%$, and $10\%$.
We use the conventional payoff values $T=5$, $R=3$, $P=1$, and $S=0$.
Each IPD lasts 400 steps, and each evaluated strategy plays 5 IPDs against each opponent. Shaded regions show one standard error calculated as $\sigma/\sqrt{5\cdot400}$, where $\sigma$ is the sample standard deviation of the average payoff per step.
}
\label{fig:eval1}
\end{figure}

The {\fontfamily{qcr}\selectfont Axelrod} library \citep{knight16} contains (as of version 4.12.0) 239 strategies for the IPD from previous literature, including all strategies mentioned so far and a zoo of different model types trained through reinforcement learning \citep{harper17}.
We use these strategies in order to evaluate CooperateISO and its two sub-strategies, as well as compare their performance with strategies which won previous tournaments or were otherwise prominently discussed in previous literature.
See Fig.~\ref{fig:eval1} for this evaluation.

Among the strategies \emph{not} introduced in the present work, we find EvolvedLookerUp2\_2\_2 to be the strongest strategy in the noise-free case, and DBS to be the strongest strategy in the presence of noise (with $\pnoise$ ranging from $1\%$ to $10\%$).
This matches the findings of \citet{harper17} (who used an older and smaller version of the {\fontfamily{qcr}\selectfont Axelrod} library).

Longterm TFT by itself is not competitive with the best-performing strategies, but outperforms the comparable strategies Generous TFT and Contrite TFT, in particular in the presence of strong noise.

CooperateISO outperforms all previous strategies at all noise levels. ISO by itself is clearly inferior to CooperateISO, in particular for zero or low noise rates, where CooperateISO is much better able to maintain cooperation. 
However, ISO outperforms its most similar strategy, DBS, despite doing simpler opponent modelling.
ISO's peformance also approaches that of CooperateISO for high noise levels, where even Longterm TFT has difficulties maintaining cooperation, and so the difference between ISO and CooperateISO becomes smaller.

\section{Desiderata for IPD strategies: beyond tournaments}\label{sec:desiderata}

Tournaments provide valuable lessons about a strategy's performance against a wide range of opponents and served as the primary benchmark for evaluating strategies in the IPD since Axelrod's work in the 1980s \citep{axelrod84}.
However, to the extend that we hope to draw wider-reaching lessons from studying strategies in the IPD, tournament performance is an imperfect measure of a strategy's benefits.
Firstly, a strategy's performance in a tournament always depends on the pool of strategies it competes against.
Secondly, it ignores aspects of the strategy that do not affect tournament performance but will become relevant in other contexts.\footnote{
As a third reason, we note that the ultimate goal in a tournament -- achieving a good rank -- is not a perfect proxy for the goal in the IPD -- achieving high payoff.
For instance, the former incentives submission of groups of strategies, in which all but one strategies seek to push the lead strategy to a high rank (by always cooperating with it and always defecting against all other participants).
Such group strategies indeed featured prominently in the 2004/2005 IPD tournaments \citep{kendall07}.
}
In particular, performance in a single tournament does not take into account the effects of other parties analyzing our strategy and adjusting their strategy to ours. 

These considerations lead us to the following three (informal) desiderata, which we motivate in more detail below. 
We stress that we are considering scenarios in which we exclusively care about maximizing our strategy's payoff.
In particular, we don't care whether we achieve higher or lower payoffs than the current opponent.\footnote{
Achieving higher payoffs than the opponent can be crucial in population games, e.g.\ when seeking to resist an invading strategy \citep{lee15}.
}
\begin{enumerate}
\item The strategy should be \emph{self-cooperating}, i.e., achieve high rates of cooperation when playing against a clone, including in the presence of significant noise. 
In particular, when playing against a clone our strategy should achieve an expected average payoff per step of $R - O(\pnoise)$ in the steady state.
So a single noise event can only lead to $O(1)$ defections in self-play, and must not lead to a longer-lasting breakdown of mutual cooperation.
\item The strategy should be \emph{cooperation-inducing}. That is, optimal (payoff-maximizing) play against our strategy should lead to an expected average payoff per step of $R - O(\pnoise)$ in the steady state for our strategy.
\item The strategy should be \emph{adaptive} (w.r.t.\ some set $\Omega$ of opponent strategies).
That is, our strategy should (be able to adapt to and thus) achieve close-to-optimal expected payoffs against all strategies from some set $\Omega$.
We ware interested in sets $\Omega$ which consist of all ``sufficiently simple'' strategies.
\end{enumerate}

Let us briefly elaborate on why we chose these particular desiderata.
\begin{itemize}
\item Not being self-cooperating limits our strategy's payoff in any context in which it has a high chance of facing a clone. This can happen, for example, if a principal deploys multiple instances of the same strategy, or if other players imitate our strategy.
\item If a strategy is not cooperation-inducing, it will do poorly as soon as an opponent is able to create a decent model of it and react to that.
TFT is the simplest possible cooperation-inducing strategy, and the historically strong performance of TFT and its generous variants lends pragmatic support to this desideratum.
\item Note that there are strategies which satisfy a strictly stronger criterion than being cooperation-inducing:
there are ``extortionate'' zero-determinant (ZD) strategies \citep{press12} which have the property that optimal (payoff-maximizing) play against them gives a payoff per step to them which is higher than $R$.
However, by construction in the IPD achieving a payoff higher than $R$ per step implies that the opponent's reward per step is lower than $R$, so such extortionate strategies cannot be self-cooperating.
\item If a strategy is not adaptive w.r.t.\ some simple opponent, it is leaving ``easy money'' on the table. 
For example, a weakness of TFT is that it always cooperates against Cooperator, while a weakness of Pavlov is that it cooperates on every second step against Defector.
At the every least, we want our strategy to be adaptive w.r.t.\ all memory-0 strategies.
That is, our strategy should always defect against all strategies which cooperate with constant probability irrespective of past actions
and thus provide no incentive to cooperate.
DBS is adaptive w.r.t.\ all memory-1 strategies\footnote{
It \emph{tries} to learn a memory-1 model of the opponent and react optimally to that. We leave aside the more difficult question of how reliably it achieves that in practice and in the presence of noise.
}, and its strong performance in tournaments lends pragmatic support to this desideratum.
\item Note that we do not add \emph{evolutionary stability} as a desideratum -- there are no evolutionary stable strategies in the IPD \citep{selten84,boyd87,farrell87,lorberbaum94}.
Instead, if the set of strategies is not restricted, evolutionary dynamics will move populations between a variety of Nash equilibria with different levels of cooperation \citep{julian18}.
\end{itemize}

Similar desiderata have been proposed by \citet{neill01}. They search for strategies which are ``self-cooperating'' (able to achieve mutual cooperation
with their clone), ``C-exploiting'' (able to exploit unconditional cooperators), and ``D-unexploitable'' (able to resist exploitation by defectors). 
Using our terminology, the latter two correspond to being adaptive w.r.t.\ $\Omega=\{\text{Cooperator}, \text{Defector}\}$, the weakest non-trivial form of adaptiveness.
We also note that being ``D-unexploitable'' is a corollary of being cooperation-inducing.

Similar desiderata have also been discussed in the context of population dynamics.
IP$_{0}$ \citep{lee15} is adaptive w.r.t.\ all memory-1 strategies (as discussed in Sec.~\ref{sec:iso}),
and self-cooperating by virtue of a noise-tolerant handshake mechanism.
These properties make IP$_{0}$ both uninvadable and a strong invader when evaluated against memory-1 strategies.
\citet{knight18} found that strategies which are best at invading populations of other strategies tend to be strategies trained to do well in tournaments;
while the best resistors of invasion invoke handshake mechanisms to cooperate with each other, but not with invaders.

\begin{figure}\centering
\includegraphics[width=1.0\textwidth]{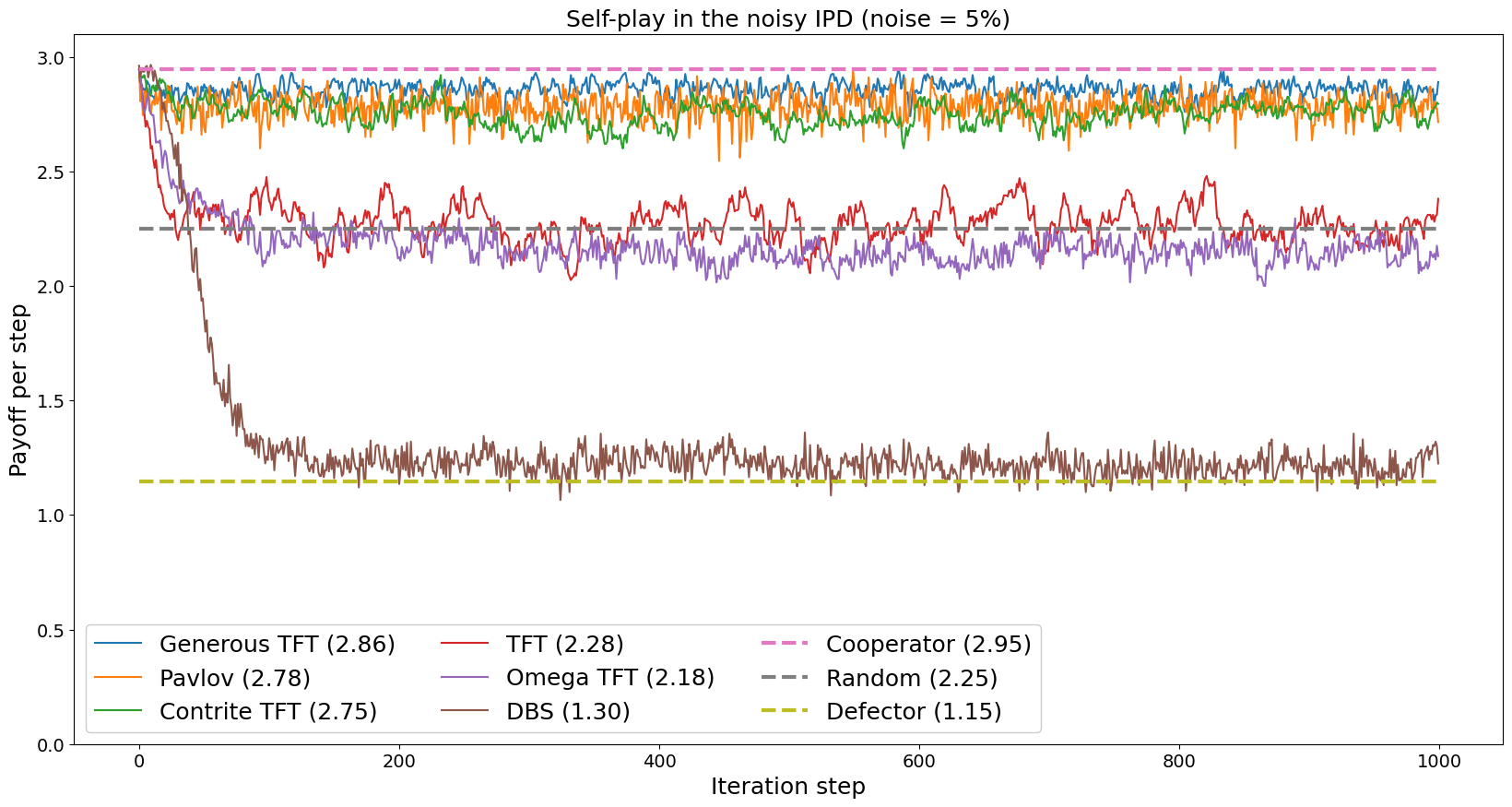}
\includegraphics[width=1.0\textwidth]{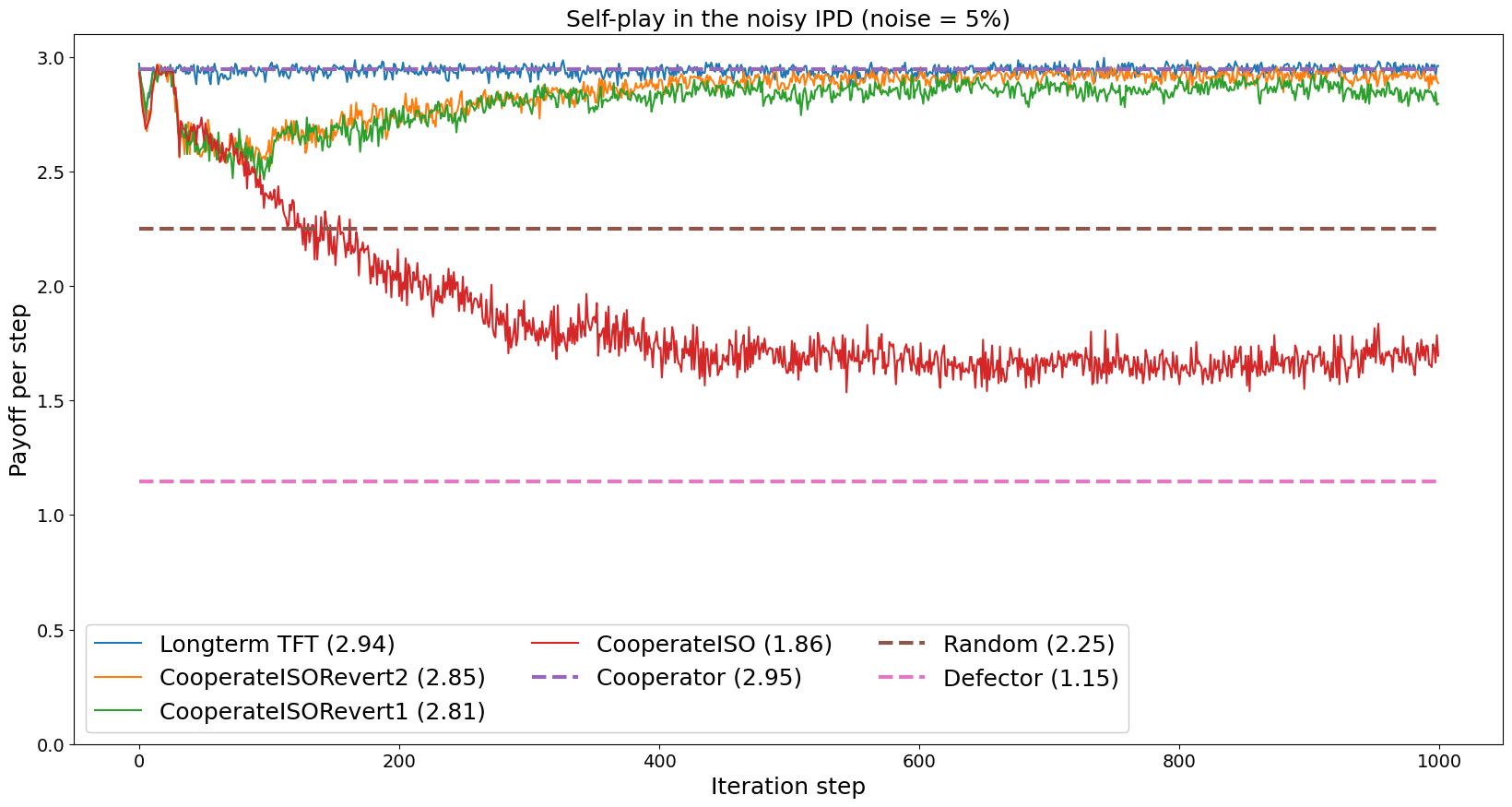}
\caption{Average payoff per step achieved by various strategies in self-play (playing against a clone) in the IPD, using $\pnoise=5\%$.
The top figure shows strategies from previous literature, the bottom figure shows strategies introduced in the present work. 
Each payoff is averaged over 100 self-play games.
The numbers in parentheses after each strategy name show the average payoff per step achieved by that strategy (averaged over 1000 steps per game and 100 games).
Dashed lines show three simple benchmarks: always cooperate, cooperate/defect randomly with $50\%$ probability, always defect, while including the effects of noise.
}
\label{fig:selfplay}
\end{figure}

Table~\ref{criteria-table} assesses fulfillment of the proposed desiderata by prominent strategies from previous literature as well as the strategies introduced in the present work.
Fig.~\ref{fig:selfplay} empirically evaluates self-cooperativeness at $\pnoise=5\%$.
CooperateISORevert1 and CooperateISORevert2 are introduced in the following section.

\begin{table}
\begin{center}
\begin{tabular}{ | c | c | c | c | }
\hline
 & Self-cooperating & Cooperation-inducing & Adaptive \\ 
\hline
Pavlov & \cmark & \xmark\footnote{In the limit of vanishing noise, always cooperating gets a payoff per step of $R$ against Pavlov while always defecting gets $(T+P)/2$. 
With the conventional IPD payoffs, both of these are equal. 
With finite noise and the conventional IPD payoffs, always defecting becomes preferable.
} & \xmark \\
TFT, Omega TFT & \xmark & \cmark & \xmark \\
Generous/Contrite TFT & \cmark & \cmark & \xmark \\
DBS  & \xmark & \xmark\footnote{The way DBS updates its opponent model makes it non-cooperation-inducing.
DBS creates a deterministic rule (of the form ``the opponent always cooperates/defects in this state'') once the opponent plays the same action 3 times consecutively in the given state.
It only abandons such a rule after 3 consecutive violations; 2 consecutive violations are forgotten if the next action follows the rule.
This makes it easy to exploit DBS: play in accordance with TFT if that is necessary to ensure that DBS' model of the opponent is TFT (and hence DBS cooperates, since that is the best response against TFT); otherwise defect (and rely on DBS ignoring up to 2 consecutive violations of its opponent model).
Empirically, we find that such a simple exploitative strategy can achieve rewards per step much higher than $R$, leading to rewards much lower than $R$ for DBS.
} & \cmark$^{(1)}$ \\
IP$_{0}$ & \cmark & \xmark & \cmark$^{(1)}$ \\
\hline
Longterm TFT & \cmark & \cmark & \xmark \\
ISO, CooperateISO  & \xmark & \xmark & \cmark$^{(1)}$ \\
CooperateISORevert1  & \cmark & \xmark & \cmark$^{(1)}$ \\
CooperateISORevert2  & \cmark & \cmark & \cmark$^{(2)}$ \\
\hline
\end{tabular}
\caption{\label{criteria-table}Fulfillment of the three proposed desiderata by prominent or newly introduced strategies.
The second half of the table describes strategies introduced in the present work.
For adaptiveness, \cmark$^{(1)}$ denotes being adaptive w.r.t.\ all memory-1 strategies, and \cmark$^{(2)}$ denotes being adaptive w.r.t.\ all \emph{non-extortionate} memory-1 strategies.}
\end{center}
\end{table}

If we accept these desiderata, a natural next question is whether it is possible for a strategy to fulfill all of them; or, more precisely, what the ``maximal'' set $\Omega$ is such that it is possible for a strategy to fulfill all of them.
We note that it is not possible for a strategy to be both cooperation-inducing and adaptive w.r.t.\ all memory-1 strategies.
This follows from the existence of extortionate ZD strategies \citep{press12}, which are memory-1 strategies and have the property that optimally responding to them (in the sense of maximizing payoffs) yields an average payoff higher than $R$ per step for them.
So if a strategy is adaptive w.r.t.\ all memory-1 strategies, optimal play against that strategy can achieve an average payoff higher than $R$ per step, implying less than $R$ per step for the strategy.
Let us define a strategy as \emph{extortionate} if the payoff-maximizing response to the strategy gives average payoffs higher than $R$ to the strategy (which implies less than $R$ for the opponent).
The best we can hope for is thus a strategy which is self-cooperating, cooperation-inducing, and adaptive to all non-extortionate memory-1 strategies. 
The following section introduces such a strategy (to the best of our knowledge, the first such strategy).

\section{Cooperate, adapt, and revert: fulfilling all desiderata}\label{sec:car}

In order to create a strategy which is self-cooperating, cooperation-inducing, and adaptive to all non-extortionate memory-1 strategies,
we start with CooperateISO and add two rules to it for reverting from the state in which it plays ISO to the state in which it plays Longterm TFT.
The basic idea is illustrated in Fig.~\ref{fig:adapt_revert}.

\subsection{Becoming self-cooperating: revert after poor performance}

CooperateISO is adaptive, but neither self-cooperating nor cooperation-inducing.
When playing against itself, it will initially be highly forgiving as long as it is in the Longterm TFT state.
Optimal play against a forgiving opponent is to defect. Both clones will thus switch to playing ISO and always defecting.
They will then update their model of the opponent to being highly defective, against which the optimal response is to defect further.
(DBS goes through a similar sequence of updates when playing against a clone.)

In order to break this cycle, we add a rule to CooperateISO to revert from the ISO state to the Longterm TFT state if the former empirically performs worse than the latter.
This ensures self-cooperativeness, and might also seem like a prudent choice when playing against other opponents. We call the resulting strategy \emph{CooperateISORevert1}.
In order to decide whether ISO's payoffs are lower than the ones of Longterm TFT, we use a standard significance test.
Formally, let $N_c$ the number of rounds for which we played Longterm TFT, $U_c$ the average payoff per step, and $\sigma_c$ the corresponding standard deviation;
and let $N_a$, $U_a$, and $\sigma_a$ be the analogous values while playing ISO.
We revert from playing ISO to playing Longterm TFT if $N_a\ge10$ and
\begin{align}\label{eq:significance}
\frac{U_c - U_a}{\sqrt{\frac{\sigma_c^2}{N_c}+\frac{\sigma_a^2}{N_a}}} > 2\ .
\end{align}

The effect of this additional rule can be seen in the bottom plot in Fig.~\ref{fig:selfplay}.
When CooperateISO plays against a clone, payoffs degrade to far below the level of mutual cooperation.
For CooperateISORevert1 (and CooperateISORevert2, which will be introduced in the following subsection), payoffs degrade similarly for the first few dozen steps.
They subsequently recover when the strategy is likely to have reverted to playing Longterm TFT.

\subsection{Becoming cooperation-inducing: revert if extorted}

CooperateISORevert1 is still not cooperation-inducing; an extortionate ZD strategy can still get rewards per step higher than $R$, if ISO responds optimally to it.\footnote{
Extortionate strategies face difficulties in the presence of noise, however; see \citet{hao15} for a discussion.
}
In order to create a strategy which is cooperation-inducing, we add a second rule for reverting from ISO to Longterm TFT:
revert to Longterm TFT if ISO is being extorted, which we define as the opponent achieving payoffs per step which are higher than $R$.
We call the resulting strategy \emph{CooperateISORevert2}.
Formally, let let $N_o$, $U_o$, and $\sigma_o$ as above describe the payoffs which the opponent achieves while playing against ISO. We revert to playing Longterm TFT if $N_o\ge10$ and $U_o-2\sigma_o/\sqrt{N_o}>R$.

By construction, it is not possible to achieve payoffs larger than $R$ in the steady state against CooperateISORevert2 while CooperateISORevert2 is in the ISO state.
TFT is the least cooperative strategy which ensures Longterm TFT's cooperation in the steady state,
and so is the best response to Longterm TFT.
The average payoff per step which TFT achieves against Longterm TFT is $R + (S+2T-3R)\pnoise+O(\pnoise^2)$,
since the two leading-order effects of noise are Longterm TFT accidentally defecting and TFT retaliating, and TFT accidentally defecting with Longterm TFT \emph{not} retaliating.
With the conventional payoffs, $S+2T-3R>0$.
TFT thus also achieves payoffs of $R + (S+2T-3R)\pnoise+O(\pnoise^2)\ge R$ against CooperateISORevert2,
making it the best response to CooperateISORevert2.
This makes CooperateISORevert2 cooperation-inducing.

\begin{figure}\centering
\includegraphics[width=1.0\textwidth]{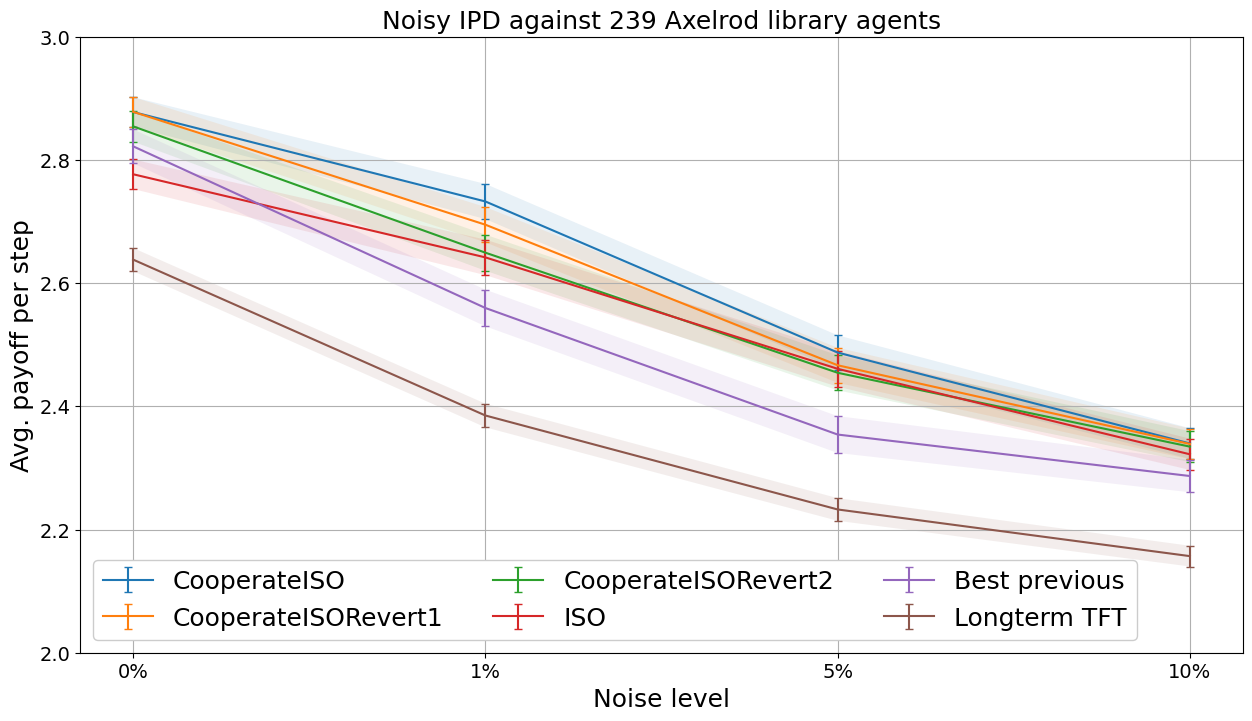}
\caption{Evaluation of all strategies introduced in the present work against the 239 strategies in the {\fontfamily{qcr}\selectfont Axelrod} library, evaluated at noise levels of $0\%$, $1\%$, $5\%$, and $10\%$.
``Best previous'' shows the performance of EvolvedLookerUp2\_2\_2 for $\pnoise=0$ and DBS for $\pnoise>0$.
Each IPD lasts 400 steps, and each evaluated strategy plays 5 IPDs against each opponent. Shaded regions show one standard error calculated as $\sigma/\sqrt{5\cdot400}$, where $\sigma$ is the sample standard deviation of the average payoff per step.
}
\label{fig:eval2}
\end{figure}

Fig.~\ref{fig:eval2} evaluates all strategies developed in the present work against the strategies in the {\fontfamily{qcr}\selectfont Axelrod} library.
We also compare all of them to the previously existing strategies which, to the best of our knowledge, show the strongest performance: EvolvedLookerUp2\_2\_2 for $\pnoise=0$ and DBS for $\pnoise>0$.
Ensuring that the strategy is both self-cooperating and cooperation-inducing, i.e., playing CooperateISORevert2 instead of CooperateISO, has a small cost in performance against this pool of opponents.
However, CooperateISORevert2 still outperforms the best-performing previously existing strategies at all noise levels.

\section{Discussion}

While even some bacteria show TFT-like behavior \citep{smith20}, the strategies introduced in the present work are arguably too complex to evolve in biological systems.
The lessons from this work are thus more relevant in the context of humans, human organizations, and human-designed systems which interact in iterated social dilemmas.
In such contexts, it is plausible that opponents imitate or analyze our strategy and adapt theirs to it. This makes it desirable that our strategy be both self-cooperating and cooperation-inducing.

When facing a diverse set of opponents, CooperateISO shows the value of enabling robust mutual cooperation, while also being ready to adapt oneself to the opponent if it does not incentivize cooperative behavior.
We have shown that such a strategy can be updated to one which is both self-cooperating and cooperation-inducing, without incurring a large loss in performance even in the more narrow context of a tournament against a fixed set of opponents.

\section*{Acknowledgements}

The author thanks Marc Harper for valuable comments on this manuscript.

\bibliographystyle{apalike}
\bibliography{bibliography}

\end{document}